\documentclass[sigconf]{acmart}  
\renewcommand\footnotetextcopyrightpermission[1]{} 
\usepackage{booktabs}

\setcopyright{none}  

\usepackage{graphicx,amsthm,amsmath,amssymb,bbm,algpseudocode,mathtools,hyperref,theoremref,amscd,amstext,fullpage,subcaption,booktabs,paralist,enumitem,todonotes,relsize,siunitx} 

\usepackage{bm,tikz,pgf,pgfplots}
\usetikzlibrary{arrows,patterns,positioning,fit,matrix,calc}
\usepackage{todonotes,marginnote, upgreek}

\usepackage{parskip}

\makeatother
\newcommand{\argmax}{\operatornamewithlimits{argmax}}
\newcommand{\argmin}{\operatornamewithlimits{argmin}}
\usepackage{algorithm,algpseudocode}
\newtheorem{defn}{Definition}
\newtheorem{lem}{Lemma}
\newcommand{\playerset}{\mathcal{I}}
\newcommand{\nplayers}{I}
\newcommand{\playeridx}{i}
\newcommand{\playeridxj}{j}
\newcommand{\action}{x}
\newcommand{\yaction}{y}
\newcommand{\zaction}{z}
\newcommand{\actionset}{X}
\newcommand{\utility}{u}
\newcommand{\reals}{{\rm I\hspace{-.07cm}R}}
\newcommand{\potential}{\Phi}

\newcommand{\timeindex}{k}
\newcommand{\timeindexl}{l}
\newcommand{\mc}[1]{\mathcal{#1}}
\newcommand{\mean}{\mu}
\newcommand{\kernel}{\kappa}
\newcommand{\obs}{y}
\newcommand{\noisevar}{{\nu}^2}
\newcommand{\noise}{\varepsilon}
\newcommand{\actionbar}{\bar{\action}}

\newcommand{\Z}{Z}
\newcommand{\GPF}{f}
\newcommand{\GPmu}{\mu}
\newcommand{\GPK}{K}

\newcommand{\TFMat}{B}
\newcommand{\Expectation}{\mathbb{E}}
\newcommand{\Cournotq}{q}
\newcommand{\CournotQ}{Q}
\newcommand{\Cournotprice}{p}
\newcommand{\Cournota}{a}
\newcommand{\Cournotb}{b}
\newcommand{\Cournotc}{c}
\newcommand{\actionsetfip}{\actionset_{\text{FIP}}}

\newcommand{\E}{\mathbb{E}}
\newcommand{\f}{f}
\newcommand{\p}{p}
\newcommand{\iteration}{\timeindex}
\newcommand{\uniquedeviator}[1]{\langle #1 \rangle}
\newcommand{\deluvec}{\Delta \mathfrak{U}}
\newcommand{\delyvec}{\Delta \mathfrak{Y}}
\newcommand{\delPhivec}{\tilde{\Phi}}
\newcommand{\phivecpath}{{\mathlarger\varphi}^\partial}
\newcommand{\stepsize}{\delta}
\newcommand{\backtracingparam}{\beta}
\newcommand{\transpose}{\intercal}
\DeclareMathOperator{\sign}{sign}
\newcommand{\eye}{I}
\newcommand{\statediffgame}{s}
\newcommand{\timediffgame}{t}
\pgfplotsset{compat=newest}
\pgfplotsset{plot coordinates/math parser=false}
\newlength\figureheight
\newlength\figurewidth
\begin{document}

\title{A Bayesian optimization approach to compute the Nash equilibria of potential games using bandit feedback}

\author{Anup Aprem}
\authornote{Anup Aprem is with the Department of Information Engineering, University of Oxford, UK. Email: aaprem@robots.ox.ac.uk}
\affiliation{}
\author{Stephen J. Roberts}
\authornote{Stephen J. Roberts is with the Department of Information Engineering \& Oxford-Man Institute of Quantitative Finance, University of Oxford, UK \& Mind Foundry Ltd. Email: sjrob@robots.ox.ac.uk}
\affiliation{}

\begin{abstract}  
  Computing Nash equilibria for strategic multi-agent systems is challenging for expensive black box systems. Motivated by the ubiquity of games involving exploitation of common resources, this paper considers the above problem for potential games. We use the Bayesian optimization framework to obtain novel algorithms to solve finite (discrete action spaces) and infinite (real interval action spaces) potential games, utilizing the structure of potential games. Numerical results illustrate the efficiency of the approach in computing the Nash equilibria of static potential games and linear Nash equilibria of dynamic potential games. 
\end{abstract}


\maketitle

\section{Introduction}
\label{sec:introduction}
Modeling strategic behavior in multi-agent systems using game theory has a rich history.  Applications of game theory include a wide range of economic phenomena such as auctions~\cite{Kri09}, oligopolies, social network formation~\cite{Jac10}, behavioral economics and political economics; just to name a few. The most common solution concept used to analyze the outcome of such a strategic interaction is the \emph{Nash equilibrium}. In a Nash equilibrium, no player benefits by deviating from their strategy~\cite{GT94}. In general, the Nash equilibrium is found as the fixed point solution of multiple single optimization problems. 

\emph{Potential games} are an important class of games first defined in~\cite{Ros73} and later popularized by~\cite{MonShap96}; refer to~\cite{Gonz16} for a recent survey. The key property of potential games is the existence of a function, called the \emph{potential function}, such that optimizing the potential function gives the Nash equilibrium. Potential games have been extensively used in the context of exploitation of common economic resource, such as in mining and fishing; see, for example, the survey~\cite{Long11}. Potential games are a natural model when the benefits that a player derives from the use of a facility is proportional to the total number of users of the same facility. An important class of potential games is that of \emph{congestion games}~\cite{VBVTF99}, widely used for understanding road transportation in urban areas~\cite{CLG16} and influence in social networks~\cite{IO11}. In the context of decision making under uncertainty and risk, \cite{Bracha12} proposed potential games as a model for interdependent preference. In addition, potential games finds use in power control in wireless network~\cite{SBP06,Hei06} and cognitive radio network~\cite{NRG04}. 

\begin{figure}[t]
  \centering
  \includegraphics[width=0.5\textwidth]{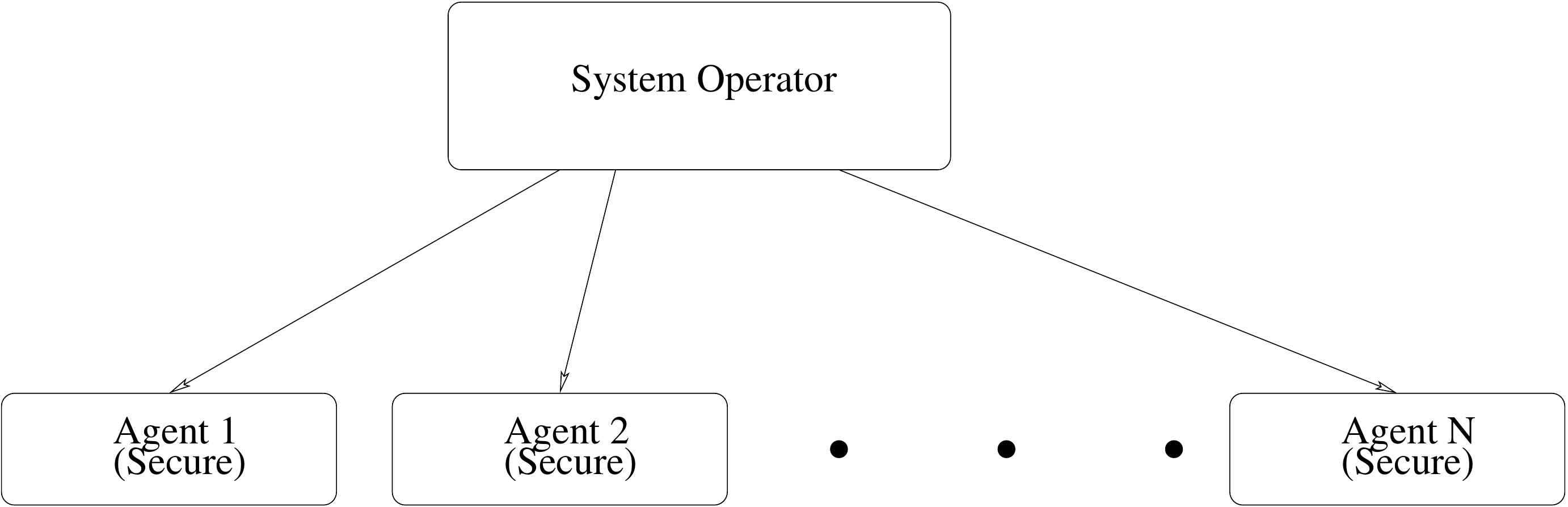}
  \caption{\footnotesize Example of a game with a black box utility function: Secure cloud computing. Each `secure' agent is unwilling to disclose its private utility function. The system operator is only able to compute the equilibria using the realized values of the private utility functions.}
  \label{fig:blk}
\end{figure}

There is an extensive literature on the techniques and algorithms for computing Nash equilibria of games, including potential games; see for example~\cite{Basar99}. However, very little is known about computing Nash equilibria of `black box' utility function or `expensive to evaluate' utility functions. \cite{Lu15}~gives the example of a secure cloud computing system as shown in Figure~\ref{fig:blk}. Each `secure' agent is unwilling to disclose its private utility function. The system operator is only able to compute the equilibrium using the realized values of the private utility functions. 
Other examples include extraction of economic resource such as water or mining where a regulator, which takes into account strategic interaction between players, have some control over the actions of the players.  
To the best of our knowledge, only~\cite{PBH18} address this problem for general games. However, several problems of exploitation of common resources, such as the cloud computing problem in~\cite{Lu15} (Figure~\ref{fig:blk}), can be modeled using potential games. Hence, in this paper, we consider the problem of computing Nash equilibria of potential games with `black box' utility functions. 

{\em Main results and Organization: } Section~\ref{sec:potential:games:intro} provides a brief introduction to potential games and their important properties. \emph{Gaussian processes}~\cite{Ras04} are a powerful, non-parametric Bayesian approach of learning and optimizing unknown functions efficiently. Section~\ref{sec:gp:intro} summarizes important results from Gaussian process regression, the framework adopted in this paper. Since potential games are characterized by a potential function, we impose a Gaussian process on the unknown potential function. The key difference between the above formulation to the standard Gaussian process regression in~\cite{Ras04} and the formulation for general games in~\cite{PBH18} is that the function (the potential in this case), on which the Gaussian process is imposed, is not directly observed.  
However, the utility of each player, which is related to the potential function, can be measured; commonly referred to as \emph{bandit feedback} in the potential game literature~\cite{HCM17}. 

This paper has the following main results:
\begin{inparaenum}[(i)]
\item Section~\ref{sec:finite:potential:games} derives an algorithm (Algorithm~\ref{algo:ne}) for efficiently computing the Nash equilibrium of potential games with finite action sets, with `black box' utility functions. 
\item Similarly, Section~\ref{sec:gp:continous:action:sets} derives an algorithm (Algorithm~\ref{algo:ne:continous:adaptive}) to compute the Nash equilibria for potential games with continuous action sets. When action sets are continuous, bandit feedback provides noisy integral of the gradients of potential function; see~\eqref{eqn:potential:continous}. Algorithm~\ref{algo:ne:continous:adaptive} presents a method which simultaneously estimates the gradient (from bandit feedback) and optimizes the unknown function using Gaussian processes. 
\item Section~\ref{sec:numerical:results} presents numerical results illustrating the efficiency of Algorithm~\ref{algo:ne} and Algorithm~\ref{algo:ne:continous:adaptive} in computing the Nash equilibrium of static games and linear Nash equilibrium of dynamic potential games, compared to existing techniques.  
\end{inparaenum}
Concluding remarks are offered in Section~\ref{sec:conclusion}.

{\em Context and Related Literature: } 
The problem considered in this paper (see, for example, Figure~\ref{fig:blk}) can be compared to the no-regret learning framework in classical game theory. 
In no-regret learning, each player, simultaneously, chooses a mixed strategy (a distribution over the action set) and obtains a pay-off associated with the strategy. The players keep a `score' based on the obtained pay-off. The mixed strategy is chosen based on the score; the most popular method being the exponential or the multiplicative weight algorithm~\cite{Tim16}. In the context of potential games~\cite{HCM17} and~\cite{CGM14} show that the exponential weight algorithm converges to the Nash equilibrium with bandit feedback. 

Inferring the unknown utilities of agents has a rich history in the revealed preference literature~\cite{Var06} in micro-economics. The revealed preference framework has been extended to the case of multiagent systems in the context of potential games in~\cite{Deb09,HKA16}. However, in comparison, this paper considers the problem of computing the Nash equilibrium, rather than characterizing the utility functions. 

{{\em Notation}: In this paper, vectors are denoted by lower case letters, e.g., $\action$, while matrices are denoted by upper case letters, e.g., $\GPK$. For a vector~$\action$, $\action^\transpose$ denotes its transpose, while for a function~$f$, $f^\prime$ denotes its derivative. The players are indexed by~$\playeridx$ and~$\playeridxj$, and~$\timeindex$ and~$\timeindexl$ denote sequence indices. For a vector $\action$, the $\playeridx^\text{th}$ component (usually, corresponding to a player) is denoted (in subscript) by $\action_\playeridx$. However, the sequence index is given in the superscript, e.g., $\action^\timeindex$. Time is denoted by~$t$. Finally, the symbol~$\sim$ is used to mean `distributed as'. }
\section{Preliminaries}
We introduce potential games in Sec.~\ref{sec:potential:games:intro} and summarize the Gaussian process regression results in Sec.~\ref{sec:gp:intro}. 
\subsection{Potential Games}
\label{sec:potential:games:intro}
Consider a game with finite number of players\footnote{In this paper, we consider potential games with only finite number of players. \emph{Population games} are potential games with infinite number of players, or rather a distribution of players. Extension of the results to population games is ongoing. }. The set of players is denoted by $\playerset = \left\{1,2,\cdots,\nplayers\right\}$. 
Player $\playeridx$ chooses an action from $\actionset_{\playeridx}$, the possible set of actions for player~$\playeridx$. 
Let $\action = \left(\action_1,\action_2,\dots,\action_\nplayers\right)$ denotes the actions of all the players. 
The utility function (or payoff) of player $\playeridx$ as a function of actions taken by all the players is given by $\utility_{\playeridx}: \actionset \mapsto \reals$, where $\reals$ denotes the real line, and $\actionset = \actionset_{1} \times \actionset_{2} \times \cdots \actionset_{\nplayers}$. 
The objective of each player is to maximize its utility function and is given by
\begin{equation}
  \action^*_{\playeridx} = \underset{\action_{\playeridx} \in \actionset_{\playeridx}}{\max\;} \utility_{\playeridx}(\action), 
  \label{eqn:utility:maximization}
\end{equation}
In the following, we will denote $\action = \left(\action_\playeridx,\action_{-\playeridx}\right)$, where $\action_{-\playeridx}$ denotes the actions of all players other than player $\playeridx$.  
A popular concept to analyze the solution of such a strategic game is the \emph{Nash equilibrium}. A solution $\action^*$ is a Nash equilibrium if 
\begin{equation}
  \action^*_{\playeridx} = \underset{\action_{\playeridx} \in \actionset_{\playeridx}}{\argmax\;} \utility_{\playeridx}(\action_\playeridx,\action^*_{-\playeridx}) \quad \forall \playeridx, 
  \label{eqn:Nash:equilibrium}
\end{equation}
i.e.\ no player benefits by deviating from $\action^*$. 

\begin{defn}[\cite{MonShap96}]
  A game is called a potential game if there exists a function~$\potential: \actionset \mapsto \reals$, called the potential function, such that
  \begin{equation}
    \utility_{\playeridx}(\yaction,\action_{-\playeridx}) - \utility_{\playeridx}(\zaction,\action_{-\playeridx}) = \potential(\yaction,\action_{-\playeridx}) - \potential(\zaction,\action_{-\playeridx}) \quad \forall \yaction,\zaction \in \actionset_{\playeridx}.
    \label{eqn:potential:defn}
  \end{equation}
\end{defn}
A \emph{finite} potential game (potential game with finite action sets) has a pure strategy Nash equilibrium~\cite{MonShap96}, i.e.\ a solution exists for~\eqref{eqn:Nash:equilibrium}. 
Another important property of potential games is the existence of a \emph{Finite Improvement Path} (FIP)~\cite{MonShap96}. A \emph{path} is a sequence of actions $\left(\action^1, \action^2,\dots,\action^\timeindex,\dots\right)$ such that, for some player~$\playeridx$, $\action^\timeindex = \left(\yaction, \action^{\timeindex-1}_{-\playeridx}\right)$; player~$\playeridx$ is referred to as the \emph{unique deviator} at $\timeindex$. Given a path, the unique deviator at $\timeindex$ can be defined as:
\begin{equation}
  \uniquedeviator{\timeindex} = \left\{ \playeridx \in \playerset: \left( \action_{-\playeridx}^\timeindex =   \action^{\timeindex-1}_{-\playeridx}\right) \& \left(\action_{\playeridx}^\timeindex  \ne \action^{\timeindex-1}_{\playeridx}\right)  \right\}. 
  \label{eqn:unique:deviator}
\end{equation}
A path is called an improvement path if $\utility_{\playeridx}(\action^\timeindex)\ge \utility_{\playeridx}(\action^{\timeindex-1}); \playeridx = \uniquedeviator{\timeindex},\;\forall \timeindex$. The FIP property for finite potential games states that every such improvement path is finite\footnote{FIP property implies that the Nash equilibrium of a potential game can be achieved by using a myopic policy in finite time.}.

\emph{Infinite Potential game}: The following lemma gives the equivalent definition of potential function for potential games with continuous action sets. 
\begin{lem}[~\cite{MonShap96}]
  \label{lem:infinite:potential:games}
  Consider a game where the action sets are intervals of real numbers. Suppose, the utility function $\utility_{\playeridx}: \actionset \mapsto \reals$ is bounded and continuously differentiable and let $\potential: \actionset \mapsto \reals$. Then, $\potential$ is a potential if and only if $\potential$ is continuously differentiable, and 
\begin{equation}
\frac{\partial \utility_\playeridx}{\partial \action_\playeridx} = \frac{\partial \potential}{\partial \action_\playeridx} \quad \forall \action. 
  \label{eqn:potential:continous}
\end{equation}
\end{lem}
Similar to the FIP property of finite games, infinite potential games have an \emph{approximate finite improvement property}~\cite{MonShap96}. 
\subsection{Gaussian Processes}
\label{sec:gp:intro}
A Gaussian process $\GPF$ indexed by $\actionset$ is a stochastic process such that for every finite collection $\action^1, \action^2, \cdots, \action^N \in \actionset$, $\bar{\GPF} = \left(\GPF(\action^1),\GPF(\action^2),\cdots,\GPF(\action^N)  \right)$ is a Gaussian random vector~\cite{Ras04}. The advantage of a Gaussian process is that it is completely specified by the mean and the covariance functions defined as follows: 
\begin{equation}
  \GPmu(\action) = \Expectation[\GPF(\action)], \quad 
    \kernel(\action, \actionbar) = \Expectation \left[ (\action -  \GPmu(\action))(\actionbar - \GPmu(\actionbar))^\transpose \right].
  \label{eqn:GP:mean:covariance}
\end{equation}
The mean and covariance function (and associated hyperparameters) encode our prior information about the model. Hence, we denote the Gaussian process by $\GPF \sim \mc{GP}\left(\GPmu,\kernel\right)$. 

Let, $\action = \left(\action^1,\action^2,\cdots,\action^N  \right)$, $\GPmu = \left(\GPmu(\action^1),\GPmu(\action^2),\cdots,\GPmu(\action^N)  \right)$, and $\GPK$ be a $N \times N$ matrix such that $\GPK_{m,n} = \kernel(\action^m, \action^n)$. It can be seen from~\eqref{eqn:GP:mean:covariance} that~$\bar{\GPF}$ is Gaussian distributed with mean~$\GPmu$ and covariance matrix~$\GPK$, i.e.\ $\bar{\GPF} \sim \mc{N}(\GPmu, \GPK)$. For any $\action^*$, the predictive distribution is Gaussian, i.e.\ $\GPF(\action^*)| \{\action^*, \bar{\GPF}, {\action} \} \sim \mc{N}(\GPmu_*, \sigma^2_*)$ with mean and variance given by: 
\begin{equation}
    \GPmu_*  	=  	\GPmu^* + {\GPK}_*^\transpose { {\GPK} }^{-1} ({\GPF} -  {\GPmu}) \quad 
    \sigma^2_* 	= 	{\GPK}_{**} - {\GPK}_*^\transpose { {\GPK} }^{-1} {\GPK}_*,
  \label{eqn:pred:gauss:dist}
\end{equation}
where, $\GPmu^* = \GPmu(\action^*)$, $\GPK_*(n) = \kernel(\action_n, \action^*)$ and $\GPK_{**} = \kernel(\action^*, \action^*)$. 
An important property is that any affine transformation of a Gaussian process is also a Gaussian process. In particular, in this paper, we are interested in differential and integral operators. Hence, the prior mean over ${\partial \GPF}/{\partial \action_i}$ is given by $\mu^D_\playeridx = {\partial \GPmu}/{\partial \action_i}$, and the covariance function is given by
\begin{equation}
  \kernel^{(D,D)}_{i,j}\left( \action , \actionbar \right) = \text{cov}\left[\frac{\partial \GPF(\action)}{\partial \action_i}, \frac{\partial \GPF(\actionbar)}{\partial \action_j}  \right] = \frac{\partial^2 \kernel}{\partial \action_i \partial \action_j}
  \label{eqn:kernel:D:D}
\end{equation}
In addition, the covariance function between the function and its derivative is given by
\begin{equation}
  \kernel^D_i\left( \action , \actionbar \right) = \text{cov}\left[{\GPF(\action)}, \frac{\partial \GPF(\actionbar)}{\partial \action_i}  \right] =  \frac{\partial \kernel}{\partial \action_i}
  \label{eqn:kernel:D}
\end{equation}
The integral operator will be introduced in Sec.~\ref{sec:gp:continous:action:sets}. 
\section{Nash Equilibria for Finite Potential Games}
\label{sec:finite:potential:games}
Since potential games are characterized by the potential function, we use a Gaussian Process model for the potential function, i.e.\ 
\begin{equation}
  \potential \sim \mc{GP}\left(\mean,\kernel \right). 
  \label{eqn:potential:GP:model}
\end{equation}
The utility function of each player is measured as below:  
\begin{equation}
  \obs_\playeridx^\timeindex = \utility_\playeridx(\action^\timeindex) + \noise_{\playeridx, \timeindex}, 
  \label{eqn:utility:noise}
\end{equation}
where $\noise_{\playeridx, \timeindex}$ is zero mean white Gaussian noise with variance $\noisevar$, i.e.\ $\noise_{\playeridx, \timeindex} \sim \mc{N}(0,\noisevar)$.  

Motivated by the FIP property, we propose the following sequential strategy for computing the Nash equilibrium. Let $\action^{\timeindex - 1}$ be the current action in the path. Consider all possible actions~$\action$ that differ only in the action of player from $\action^{\timeindex - 1}$ i.e.\ $\action \in \actionset_{\text{FIP}}^{\timeindex - 1} = \left\{ \action: \left( \action_{-\playeridx} =   \action_{-\playeridx}^{\timeindex - 1}\right) \& \left(\action_{\playeridx}  \ne \action_{\playeridx}^{\timeindex - 1}\right) \right\}$. 
For computing the next action in the path, we would have chosen an action that improves the utility of player $\uniquedeviator{\action, \action^{\timeindex -1}}$, where with an abuse of notation $\uniquedeviator{\action, \action^{\timeindex -1}}$ denotes the unique deviator between $\action$ and $\action^{\timeindex -1}$ as in~\eqref{eqn:unique:deviator}. However, since the unknown potential function (and hence the utility function) is modeled using a random function, we will use the following \emph{one-step lookahead} criterion~\cite{OGR09}. The one-step lookahead criterion automatically promotes trade-off between exploration and exploitation; see discussion in~\cite{OGR09}. The one-step lookahead criterion is very similar to the classical expected improvement criterion~\cite{Jones98}.  

The one-step lookahead criterion selecting the action that maximizes the following `utility function'
\begin{equation}
  \eta(\action) = \underset{{\action \in \actionsetfip^{\timeindex -1}}: {\uniquedeviator{\action, \action^{\timeindex -1}} = \playeridx}}{\argmax\;}{\E\left[\utility_\playeridx(\action) -  \utility_\playeridx(\action^{\timeindex - 1})  \right]_{+}}, 
  \label{eqn:one:step:lookahead}
\end{equation}
where, $\left[ x \right]_+ = \max\left\{x,0  \right\}$. 
However, from the definition of potential games in~\eqref{eqn:potential:defn}, 
\begin{equation*}
  	\begin{aligned}
	  \utility_{\playeridx}(\action) -  \utility_{\playeridx}(\action^{\timeindex - 1})   &=  \potential(\action) -  \potential(\action^{\timeindex - 1})  .
	\end{aligned}
\label{eqn:one:step:lookahead:potential}
\end{equation*}
Define, $\Z = \potential(\action) -  \potential(\action^{\timeindex - 1})$ and 
\begin{equation}
  \begin{aligned}
    \delPhivec &= \left(\potential(\action^1),\potential(\action^2),\cdots,\potential(\action^{\timeindex - 1}), \potential(\action^\timeindex = \action) \right), \\
    \deluvec &= \left(\Delta \utility_{\uniquedeviator{1}}, \Delta \utility_{\uniquedeviator{2}}, \cdots  ,\Delta \utility_{\uniquedeviator{\timeindex-1}}\right), \\ 
    \delyvec &= \left(\Delta \obs_{\uniquedeviator{1}}, \Delta \obs_{\uniquedeviator{2}}, \cdots  ,\Delta \obs_{\uniquedeviator{\timeindex-1}}\right), 
\end{aligned}
  \label{eqn:diff:delu}
\end{equation}
where, $\Delta \utility_{\uniquedeviator{\timeindex}} = \utility_{\uniquedeviator{\timeindex}}^\timeindex - \utility_{\uniquedeviator{\timeindex}}^{\timeindex-1}$, and  $\Delta \obs_{\uniquedeviator{\timeindex}} = \obs_{\uniquedeviator{\timeindex}}^\timeindex - \obs_{\uniquedeviator{\timeindex}}^{\timeindex-1}$. 
Since the potential function is modeled as a Gaussian process in~\eqref{eqn:potential:GP:model}, $\delPhivec$ is Gaussian, i.e. $\delPhivec \sim \mc{N}(\GPmu, \GPK)$, with mean  
$\GPmu(\timeindex) = \GPmu(\action^\timeindex)$, and covariance $\GPK(\timeindex, \timeindexl) = \kernel(\action^\timeindex, \action^{\timeindexl})$. 
Hence,
\begin{equation}
  \TFMat \delPhivec = \begin{pmatrix}\deluvec \\ \Z \end{pmatrix} \sim \mc{N}\left(\bar{\mean} = \TFMat \mean, \bar{\GPK} = \TFMat \GPK \TFMat^\transpose\right),
  \label{eqn:diffGP}
\end{equation}
where, $\TFMat$ is the differencing matrix given by
\begin{equation*}
  \TFMat({i,j}) =
  \begin{cases}
    -1 & j=i \\
    1  & j=i+1 \\
    0  & \text{else}.
  \end{cases}
  \label{eqn:transform:matrix}
\end{equation*}
Consider the partition of the mean and the covariance matrix in~\eqref{eqn:diffGP} as below: 
\begin{equation}
    \bar{\mu} = \left[  \begin{array}{c|c} \underset{(\timeindex-1)}{\bar{\mu}_1} & \underset{1}{\bar{\mu}_2}  \end{array} \right], \;
    \bar{K}   = \left[ \begin{array}{c|c} \underset{(\timeindex-1)\times (\timeindex-1)}{\bar{K}_{1,1}} & \underset{(\timeindex-1)\times 1}{\bar{K}_{1,2}} \\ \hline \underset{1\times (\timeindex-1)}{\bar{K}_{2,1}} & \underset{1\times 1}{\bar{K}_{2,2}} \end{array} \right]. 
  \label{eqn:mean:covariance:partition}
\end{equation}
The posterior distribution of $\Z$ given the observations~$\delyvec$ in~\eqref{eqn:diff:delu} can be obtained using~\eqref{eqn:pred:gauss:dist} as Gaussian with mean and variance given by 
\begin{equation}
  \begin{aligned}
    \mu_\Z  		&=  	\bar{\mu}_2 + \bar{\GPK}_{2,1} {\left( \bar{K}_{1,1}  + 2\noisevar \eye \right)}^{-1} \left( \delyvec -  {\bar{\mu}_1} \right), \\
    \sigma^2_\Z 	&= 	\bar{\GPK}_{2,2} - {\bar{K}_{2,1}} {\left( \bar{K}_{1,1}  + 2\noisevar \eye \right)}^{-1} {\bar{K}_{1,2}},
  \end{aligned}
  \label{eqn:pred:gauss:dist:Z}
\end{equation}
where, we have used the fact that $\Delta \obs_{\uniquedeviator{\timeindex}}\bigr|_{\Delta \utility_{\uniquedeviator{\timeindex}}}$, being the difference of two Gaussian variables has variance $2\noisevar$; see~\eqref{eqn:utility:noise}. 
Given the mean and variance in~\eqref{eqn:pred:gauss:dist:Z}, it can be shown that~\cite{OGR09}
\begin{equation}
  {\E\left[\Z \right]_{+}} =  \mu_\Z\left( 1 - \upphi\left( {-\mu_\Z}/{\sigma_\Z} \right) \right) + \frac{\sigma_\Z}{\sqrt{2\pi}} \exp{\left(-{\mu^2_\Z}/{2\sigma^2_\Z} \right)} ,
  \label{eqn:Z:plus:dist}
\end{equation}
where, $\upphi$ is the standard Gaussian distribution. 

Algorithm~\ref{algo:ne} summarizes the Bayesian optimization approach to compute Nash equilibria in finite potential games. 
\begin{algorithm}
\caption{Bayesian algorithm to compute Nash equilibrium in finite potential games}
\label{algo:ne}
\begin{algorithmic}[1]
       \State Choose random initial action $\action^0$. 
       \For{$\timeindex=1,2,\dots$: }
       \State Construct $\actionset_{\text{FIP}}^{\timeindex - 1}= \left\{ \action: \left( \action_{-\playeridx} =   \action_{-\playeridx}^{\timeindex - 1}\right) \& \left(\action_{\playeridx}  \ne \action_{\playeridx}^{\timeindex - 1}\right) \right\}$. 
       \State Compute $\delyvec$ according to~\eqref{eqn:diff:delu}. 
       \For{$\action \in \actionset_{\text{FIP}}^{\timeindex - 1}$} 
       \State Compute mean and covariance of $\Z$ using the partitions from~\eqref{eqn:mean:covariance:partition} according to~\eqref{eqn:pred:gauss:dist:Z}. 
       \State Compute ${\E\left[\Z \right]_{+}}$ according to~\eqref{eqn:Z:plus:dist}. 
       \EndFor
       \State Choose  $\action^{\timeindex} = \underset{\action \in \actionsetfip^{\timeindex -1}}{\argmax \;} {\E\left[\Z \right]_{+}}$
       \EndFor
\end{algorithmic}
\end{algorithm}
\section{Nash Equilibria for Infinite Potential Games}
\label{sec:gp:continous:action:sets}
This section considers potential games with continuous action sets. 
When action sets are continuous, the potential function of the game is related to the utility function of the players through the derivative as in~\eqref{eqn:potential:continous}. Similar to finite potential games, we propose generating a path. Let $\left(\action^1, \action^2,\cdots ,\action^{\timeindex - 1}\right)$ be the current state of the path.  
Let the next action in the path be $\action^\timeindex = \left(\yaction, \action^{\timeindex-1}_{-\playeridx}\right)$. We propose to update $\yaction$ as below: 
\begin{equation}
  \yaction = \action^{\timeindex - 1}_\playeridx + \delta_\playeridx \; \frac{\partial \potential}{\partial \action_\playeridx} \Bigr|_{\action = \action^{\timeindex - 1}}. 
  \label{eqn:gradient:ascent}
\end{equation}
The gradient algorithm in~\eqref{eqn:gradient:ascent} requires the following: \begin{inparaenum}[(i)] 
\item estimating the gradient $\frac{\partial \potential}{\partial \action_\playeridx}$, 
\item choosing appropriate step-size $\delta_\playeridx $, and,
\item choosing the player~$\playeridx$ to update. 
\end{inparaenum}
We deal with each of these steps in turn in Sec.~\ref{subsec:estiamte:grad:potential}, Sec.~\ref{sec:adaptive:step:size:algo}, and Sec.~\ref{subsec:player:update}, respectively. 
\subsection{Estimating the potential gradient}
\label{subsec:estiamte:grad:potential}
Let, $x^k = \left( y, x^{k-1}_{-i} \right)$ and $\Delta u_i^k = u_i^k - u_i^{k-1}$, where $\playeridx$~is the unique deviator from $\timeindex -1$ to $\timeindex$.  Consider the path:
\begin{equation}
  r^k(\tau) = x^{k-1} + \tau\left( x^k - x^{k-1} \right) \; \tau \in \left[ 0,1 \right].
  \label{eqn:integral:path}
\end{equation} 
Then, 
\begin{equation}
  \int_{r^k(\tau)}  \frac{\partial \potential}{\partial \tau} d\tau = \potential(x^k) -  \potential(x^{k-1}) = \Delta u_i^k,
  \label{eqn:line:integral:potential:diff}
\end{equation}
where the first equality in~\eqref{eqn:line:integral:potential:diff} follows from the fact that the line integral of a scalar field, such as the potential, depends only on the end points. The second equality in~\eqref{eqn:line:integral:potential:diff} follows from the definition of the potential games in~\eqref{eqn:potential:defn}. The integral in~\eqref{eqn:line:integral:potential:diff} can be re-written as follows: 
\begin{align}
  & \int_{r^k(\tau)}  \frac{\partial \potential}{\partial \tau} d\tau =  \int_{0}^1 \sum_j \frac{\partial \potential}{\partial r_j} \frac{d r_j}{d \tau} d\tau =  \int_{0}^1 \sum_j \frac{\partial \potential}{\partial r_j} \left( x_j^{k} - x_j^{k-1} \right) d\tau  \nonumber\\
  &=  \int_{0}^1 \frac{\partial \potential}{\partial r_i} \left( x_i^{k} - x_i^{k-1} \right) d\tau =  \left( x_i^{k} - x_i^{k-1} \right) \int_{0}^{1} \frac{\partial \potential}{\partial x_i} d\tau, \label{eqn:path:integral:potential}
\end{align}
where the first equality is obtained by `total derivative' formulae, and the last equality is due to $\playeridx$ being the unique deviator from $\action^{\timeindex - 1}$ to $\action^\timeindex$. Hence, from~\eqref{eqn:line:integral:potential:diff} and~\eqref{eqn:path:integral:potential}, the change in utility $\Delta u_i^k$ can be interpreted as integral observations of the differential of the potential along the $x_i$ direction. 

Define, 
\begin{equation}
  \begin{aligned}
    \Lambda &= \text{diag} \left( \Delta x_{<1>}, \Delta x_{<2>}, \cdots  ,\Delta x_{<\timeindex-1>}   \right), \\
    \phivecpath &= \left( \frac{\partial \potential}{\partial x_{<1>}}, \frac{\partial \potential}{\partial x_{<2>}}, \cdots \frac{\partial \potential}{\partial x_{<\timeindex-1>}} \right).
  \end{aligned}
  \label{eqn:potential:vectors}
\end{equation}
Using the vectors in~\eqref{eqn:potential:vectors}, along the path, \eqref{eqn:path:integral:potential} can be written in vector notation as 
\begin{equation*}
  \deluvec = \Lambda \left( \int \odot \; \phivecpath \right),
  \label{eqn:line:integral:vector}
\end{equation*}
where, $\deluvec$ is as in~\eqref{eqn:diff:delu} and $\odot$ is the Hadamard (point-wise) product. $\deluvec$ is a Gaussian process since it is obtained by linear operators on the gradient of the potential function which is a Gaussian process. Let the potential gradient vector be denoted by $\Phi^\partial = \left(\frac{\partial \potential}{\partial x_{1}}, \frac{\partial \potential}{\partial x_{2}}, \cdots \frac{\partial \potential}{\partial x_{\nplayers}}\right)$. The joint Gaussian process can be shown to be equal to
\begin{equation}
  \begin{pmatrix}
   \Phi^\partial  \\ \deluvec
  \end{pmatrix}
  \sim \mc{GP}\left(
  \begin{pmatrix}
    \mu^D \\ \mu^{\deluvec}
  \end{pmatrix}
  ,
  \begin{pmatrix}
    \kernel^{(D,D)} & \gamma^T \\ \gamma & \pi
  \end{pmatrix}
\right),
  \label{eqn:joint:gp:potential:delu}
\end{equation}
where,
\begin{align}
  \mu^{\deluvec}_{\timeindex} &= \Delta x_{\uniquedeviator{\timeindex}} \int \mu^D_{\uniquedeviator{\timeindex}}\left(r^{\timeindex}(\tau)\right) \; d\tau \nonumber\\
    \gamma_{\timeindex,\playeridx}(\action) &=  \Delta x_{\uniquedeviator{\timeindex}} \int_0^1 \kernel^{(D,D)}_{\uniquedeviator{\timeindex}, \playeridx}\left( r^{\timeindex}(\tau), \action \right) \; d\tau \label{eqn:potential:covariance} \\ 
    \pi_{\timeindex,\timeindexl} &= \Delta x_{\uniquedeviator{\timeindex}} \Delta x_{\uniquedeviator{\timeindexl}} \int_0^1 \int_0^1 \kernel^{(D,D)}_{{\uniquedeviator{\timeindex}}, {\uniquedeviator{\timeindexl}}}\left( r^{\timeindex}(\tau), r^{\timeindexl}(\bar{\tau})\right) \; d\tau \; d\bar{\tau}, \nonumber
\end{align}
where, $r^{\timeindex}(\tau)$ is the path in~\eqref{eqn:integral:path}, $\Delta \action_{\uniquedeviator{\timeindex}} $ is as in~\eqref{eqn:diff:delu}, and $\mu^D$ and $\kernel^{(D,D)}$ are the mean and covariance function of the Gaussian process $\Phi^\partial$ described in Sec.~\ref{sec:gp:intro}.

The posterior distribution process of the potential gradient can be obtained using~\eqref{eqn:pred:gauss:dist} as 
\begin{equation}
  \begin{aligned}
    \mu^D\Bigr|_{\delyvec} &= \mu^D + \gamma^\transpose {\left( {\mathlarger \pi} + 2\noisevar \eye \right)}^{-1} \left(\Delta u - \mu^{\deluvec} \right), \\
    \kernel^{(D,D)}\Bigr|_{\delyvec} &=  \kernel^{(D,D)} - \gamma^\transpose {\left( {\mathlarger \pi} + 2\noisevar \eye \right)}^{-1}\gamma,
  \end{aligned}
  \label{eqn:posterior:potential:diff}
\end{equation}
where, ${\delyvec}$ is as in~\eqref{eqn:diff:delu} and we have again used the fact that $\Delta \obs_{\uniquedeviator{\timeindex}}\bigr|_{\Delta \utility_{\uniquedeviator{\timeindex}}}$, being the difference of two Gaussian variables, has variance $2\noisevar$. Having estimated the potential gradient, the next section considers the problem of selecting the step size. 
\subsection{Choosing the step size}
\label{sec:adaptive:step:size:algo}
The \emph{ideal} step size for a gradient algorithm such as the one in~\eqref{eqn:gradient:ascent} is given by the second derivative (or the Hessian). 
However, obtaining the second derivatives is computationally taxing. Hence, a practical strategy is to perform a \emph{line search}. A line search algorithm tries a sequence of candidate value of step size, and stops to accept one of the candidate values when the conditions are `acceptable'. A popular method of line search is given by the \emph{backtracing} algorithm. The key idea of backtracing is to start with a maximum step size, i.e.\ $\stepsize_0 = \stepsize_{\text{max}}$ and then iteratively generate step sizes as follows: $\stepsize_m = \backtracingparam \stepsize_{m - 1}$, $\backtracingparam \in (0,1)$ being the backtracing parameter, until an `acceptable' step size is found. The following conditions popularly known as Wolfe conditions~\cite{NocWri06} provide step size that ensures convergence of a gradient algorithm. 
\begin{defn}[Wolfe's conditions]
  \label{defn:wolfe:conditions}
  Given a function $\f$ and an ascent direction $\p$, i.e.\ $\p^\transpose {\f^\prime(\action^\iteration)} > 0$, with $0 < c_1 < c_2 \le 1$, a step-size $\delta$ is acceptable if: 
  \begin{itemize}
    \item \textbf{Sufficient Increase}: $\f(\action^\iteration + \delta \p) - \f(\action^\iteration) \ge c_1 \delta {\f^\prime(\action^\iteration)}^\transpose \p$
    \item \textbf{Curvature Condition}: $c_2 {\f^\prime(\action^\iteration)}^\transpose \p \ge {\f^\prime(\action^\iteration + \delta \p)}^\transpose \p$
  \end{itemize}
  \label{def:wolfe}
\end{defn}
The first condition ensures that there is `sufficient' increase in function value in the direction $\p$. The second condition ensures that the slope of the function decreases. Typically the values are chosen as: $c_1 = \num{1e-4}$ and $c_2 = 0.8$~\cite{NocWri06}. In the Bayesian optimization framework, \cite{MH15}~derives a probabilistic approach to the Wolfe conditions. The key idea in~\cite{MH15} is to compute the probability that the Wolfe conditions in Defn.~\ref{def:wolfe} are satisfied and then only consider candidates that pass a threshold, denoted by $c_w$. 

\textbf{Computing the probability of Wolfe conditions}: 
From Defn.~\ref{defn:wolfe:conditions}, computing the probability of Wolfe conditions require modelling the function along with its derivatives. In addition, we need to choose the ascent direction $\p$. Choosing to update player $\playeridx$ in~\eqref{eqn:gradient:ascent}, we choose the ascent direction as below: 
\begin{equation}
  p_\playeridx = \sign{\left( \E \left( \frac{\partial \potential}{\partial \action_\playeridx} \right) \right)} e_i,
  \label{eqn:ascent:direction}
\end{equation}
where, $e_i$ is the standard basis vector. The posterior distribution of ${\partial \potential}/{\partial \action_\playeridx}$ is obtained from~\eqref{eqn:posterior:potential:diff} in Sec.~\ref{subsec:estiamte:grad:potential}. 

Let the step size be~$\delta$. For player~$\playeridx$, define $\bar{\action}^\timeindex = \action^\timeindex +  \delta \p_\playeridx $, the next possible action in the path, where the ascent direction $\p_\playeridx$ is given by~\eqref{eqn:ascent:direction}. Let~$\Phi^\playeridx$ be the vector of potential value at~$\bar{\action}^\timeindex$ and $\action^\timeindex$, i.e.\ $\Phi^\playeridx = \potential (\bar{\action}^\timeindex) ,  \potential ( \action^\timeindex ))$, and, $\Phi^{\partial, \playeridx}$ be the gradient of $\Phi^{\playeridx}$ with respect to $\action_\playeridx$, i.e.\ $\Phi^{\partial, \playeridx} = (\partial \potential(\bar{\action}^\timeindex)/\partial \action_\playeridx, \partial\potential( \action^\timeindex )/\partial \action_\playeridx)$. 
The joint Gaussian process of~$\Phi^\playeridx$, $\Phi^{\partial, \playeridx}$ and the observations~$\deluvec$ in~\eqref{eqn:diff:delu} can be shown to be equal to 
\begin{equation}
  \begin{pmatrix}
    \Phi^i \\ \Phi^{\partial,i} \\ \deluvec
  \end{pmatrix}
  \sim \mc{GP}\left(
  \begin{pmatrix}
    \mu \\ \mu^D \\ \mu^{\deluvec}
  \end{pmatrix}
  ,
  \begin{pmatrix}
    \kernel & \kernel^D & \eta^T \\
    \kernel^D & \kernel^{(D,D)} & \gamma^T \\ 
    \eta & \gamma & \pi
  \end{pmatrix}
\right),
  \label{eqn:joint:gp:gpdiff:potential:delu}
\end{equation}
where, $\mu^{\deluvec}$, $\gamma$ and $\pi$ are as in~\eqref{eqn:potential:covariance}. $\kernel^D$ is the covariance between the observation and its derivative, see Sec.~\ref{sec:gp:intro}. Using the kernel~$\kernel^D$, $\eta$ in~\eqref{eqn:joint:gp:gpdiff:potential:delu} is 
\begin{equation}
  \eta_\timeindex = \Delta \action_{\langle \timeindex \rangle} \int_0^1 \kernel^{D}_{\langle \timeindex \rangle}\left( r^{\timeindex}(\tau), \action \right) \; d\tau,
  \label{eqn:eta:covariance}
\end{equation}
where, $r^{\timeindex}(\tau)$ is the path defined in~\eqref{eqn:path:integral:potential}. 
The posterior distribution of $\left( \Phi^i, \Phi^{\partial,i} \right) \sim \mc{GP}\left(\mu^\delta,\kernel^\delta\right)$ can be obtained similar to~\eqref{eqn:posterior:potential:diff}, as follows
\begin{equation}
  \begin{aligned}
    \mu^\delta\Bigr|_{\delyvec} &= \begin{pmatrix} \mu \\  \mu^D \end{pmatrix} + \begin{pmatrix} \eta \\ \gamma \end{pmatrix}^\transpose {\left( {\mathlarger \pi} + 2\noisevar \eye \right)}^{-1} \left(\delyvec - \mu^{\deluvec} \right), \\
    \kernel^\delta\Bigr|_{\delyvec} &=  \kernel^{(D,D)} - \begin{pmatrix} \eta \\ \gamma \end{pmatrix}^\transpose {\left( {\mathlarger \pi} + 2\noisevar \eye \right)}^{-1} \begin{pmatrix} \eta \\ \gamma \end{pmatrix},
  \end{aligned}
  \label{eqn:posterior:diff:forwolfe}
\end{equation}
Similar to~\eqref{eqn:posterior:potential:diff}, we have used the fact that $\Delta \obs_{\uniquedeviator{\timeindex}}\bigr|_{\Delta \utility_{\uniquedeviator{\timeindex}}}$, being the difference of two Gaussian variables, has variance $2\noisevar$.  

The Wolfe conditions in Defn.~\ref{defn:wolfe:conditions} in vector notation is as follows: 
\begin{equation}
  \begin{pmatrix} a_\iteration \\ b_\iteration \end{pmatrix} 
  = \begin{pmatrix} -c_1 & 0 & -1 & 1 \\ 0 & 0 & c_2 & -1 \end{pmatrix}   
  \begin{pmatrix} \Phi^i \\ \Phi^{\partial,i} \end{pmatrix}.
  \label{eqn:pwolfe}
\end{equation}
Then, $a_\iteration$ and $b_\iteration$ in~\eqref{eqn:pwolfe} are jointly Gaussian. The probability that the Wolfe conditions are satisfied is given by $P_w = P( (a_\iteration \ge 0) \& (b_\iteration \ge 0) )$. Several packages, such as the one in~\cite{DW90}, are available to compute this probability efficiently. 

\subsection{Choosing the player to update}
\label{subsec:player:update}
Section~\ref{subsec:estiamte:grad:potential} considered the problem of estimating the gradient of the potential function and Sec.~\ref{sec:adaptive:step:size:algo} considered the problem of selecting an `acceptable' step size $\stepsize_\playeridx$ (probability of Wolfe condition above the $c_w$ threshold), for each player $\playeridx$. 

In this section, we consider the problem of choosing which player to update. As in the finite potential games, we select the player which provides the maximum `expected improvement' in the one-step lookahead criterion~\eqref{eqn:one:step:lookahead}, i.e.\ 
\begin{equation}
\playeridx^* = \underset{\playeridx}{\argmax\;}{\E\left[\potential ( \action^\timeindex +  \delta_\playeridx \p_\playeridx ) -  \potential ( \action^\timeindex )\right]_{+}}
  \label{eqn:choose:player:infinite}
\end{equation}
It is straightforward to compute~\eqref{eqn:choose:player:infinite} using the posterior probability in~\eqref{eqn:posterior:diff:forwolfe} and the formulae in~\eqref{eqn:Z:plus:dist}. 

Algorithm~\ref{algo:ne:continous:adaptive} summarizes the various steps explained above to compute the Nash equilibrium for infinite potential games. 
\begin{algorithm}
\caption{Bayesian algorithm to compute Nash equilibria in infinite potential games}
\label{algo:ne:continous:adaptive}
\begin{algorithmic}[1]
  \Require Line search parameters $c_1$ and $c_2$, Maximum step size $\stepsize_\text{max}$, Wolfe probability threshold $c_{w}$, 
       \State Choose random initial action $\action^0$. 
       \For{$\timeindex=1,2,\dots$: }
       \State Estimate gradients using~\eqref{eqn:posterior:potential:diff}.
       \For{player $\playeridx = 1,2,\cdots, n$}
       \State Choose ascent direction $\p_\playeridx$ using~\eqref{eqn:ascent:direction}. 
       \State $\stepsize_\playeridx = \stepsize_\text{max}$. 
       \Repeat
       \State Compute Wolfe probability~$P_w$ using~\eqref{eqn:posterior:diff:forwolfe}--\eqref{eqn:pwolfe}. 
       \State Reduce step size $\stepsize_\playeridx = \beta \stepsize_\playeridx$
       \Until \{ $P_w \ge c_w$ \}
       \EndFor
       \State Choose player $\playeridx^*$ according to~\eqref{eqn:choose:player:infinite}. 
       \State Update $\action^\timeindex$ according to~\eqref{eqn:gradient:ascent} with step-size $\stepsize_{\playeridx^*}$
       \EndFor
\end{algorithmic}
\end{algorithm}
\section{Numerical Results}
\label{sec:numerical:results}
In this section, we apply the methods in Sec.~\ref{sec:finite:potential:games} and Sec.~\ref{sec:gp:continous:action:sets} to compute the Nash equilibrium of static potential games and the linear Nash equilibrium of dynamic potential games. In this numerical section, we will use the following choice of mean and covariance function:
\begin{equation}
  \GPmu(\action) = 0, \quad \kernel_{\text{se}} = \ell^2 \exp\left[- \frac{1}{2} \sum_{j = 1}^\nplayers  \frac{\left( \action_j - \bar{\action}_j \right)^2}{\lambda_j^2}\right], 
  \label{eqn:mu:kernel:choice}
\end{equation}
where, the hyperparameters $\lambda_j$, known as the input scale length, determines the relevance of each dimension. The hyperparameter $\ell$ controls the magnitude of the output. Setting the mean function to zero is motivated by the ordinal nature\footnote{An ordinal utility function is unique up to increasing monotone transformation. For example, adding a positive constant term to the utility function of the players does not change the solution in~\eqref{eqn:utility:maximization}.} of the utility function of the players (and, hence the potential function). The choice of the squared exponential kernel $\kernel_\text{se}$ is motivated by 
\begin{inparaenum}[(i)]
	\item Lemma~\ref{lem:infinite:potential:games} requires that the potential function be continuously differentiable, 
	\item existence of analytical expressions for the kernels in~\eqref{eqn:kernel:D:D} and~\eqref{eqn:kernel:D}.
\end{inparaenum} 
In particular, for the squared exponential kernel: 
\begin{equation*}
  \begin{gathered}
    \kernel^{(D)}_i =  \kernel_{\text{se}} \left(  \frac{\left( \action - \actionbar \right)_i }{\lambda_i^2} \right),\\
    \kernel^{(D,D)}_{i,j} =  \kernel_{\text{se}} \left( \frac{\delta_{i,j}}{\lambda_i^2} + \frac{\left( \action - \actionbar \right)_i \left( \actionbar - \action \right)_j}{\lambda_i^2 \lambda_j^2} \right), 
  \end{gathered}
  \label{eqn:exponential:kernel:derivatives}
\end{equation*}
where, $\delta_{i,j}$ is the Kronecker delta function.  
\subsection{Static Potential Games: Cournot oligopoly}
\label{sec:numerical:results:static:games}
Cournot oligopoly is a market with $\nplayers$~players for a single good, where each player manufactures a certain quantity of good given by $\Cournotq_\playeridx$. The price is dictated by the market and is a function of the total quantity produced by all the players, given by $\Cournotprice = \Cournota - \Cournotb \CournotQ$, where $\CournotQ = \sum_{\playeridx} \Cournotq_\playeridx$ and $\Cournota, \Cournotb > 0$. Each player has a cost function given by $\Cournotc_\playeridx(\Cournotq_\playeridx)$, a function of the quantity produced by each player. The Cournot oligopoly is known to be a potential game~\cite{MonShap96}. 

To illustrate the main results, we consider a version of the problem with $2$ players, with the following parameters: 
\begin{align}
  &\Cournotc_\playeridx(\Cournotq_\playeridx)=d_\playeridx {\Cournotq_\playeridx}^{\beta_\playeridx} \quad \utility_\playeridx = \Cournotq_\playeridx \Cournotprice - \Cournotc_\playeridx(\Cournotq_\playeridx) \quad \playeridx=1,2 \quad \beta_\playeridx \in \left( 0,2 \right), \nonumber\\
  &a = 10, \quad b = 1, \quad  d_\playeridx = 5, \quad  \beta_1 = 0.95, \quad \beta_2 = 1.95, \label{eqn:params:cournot}
\end{align}
where, we have used the exponential cost function for the players in~\eqref{eqn:params:cournot}. The parameter $\beta_\playeridx$ characterizes the rate of growth of cost~\cite{Wal63}. The first player models an agent with low rate of cost growth ($\beta_1 < 1$) i.e.\ the agent profits from more production. However, the second player models an agent with high rate of cost growth ($\beta_2 > 1$), i.e.\ the payoff decreases with more production. Each agent sells quantity $\Cournotq_\playeridx$ at price $\Cournotprice$. The utility of the player is the profit given in~\eqref{eqn:params:cournot}. 

As can be seen that the price goes to zero when $Q=10$. Hence, we restrict our search space from $(0,10]$. Due to the nature of the cost function in~\eqref{eqn:params:cournot}, straight-forward analytic methods cannot be used for computing the Nash equilibrium, even for the case of a~$2$ player game.  

\textbf{Finite Game Formulation}: 
The action space of the Cournot problem is continuous. To apply the finite game formulation in Sec.~\ref{sec:finite:potential:games}, we discretize the action space into a $31 \times 31$ grid. Algorithm~\ref{algo:ne} is run with the following parameters:
\begin{equation}
  \lambda_\playeridx^2 = \sqrt{30}, \quad \ell = 1. 
\label{eqn:params:cournot:algo:finite}
\end{equation}
Algorithm~\ref{algo:ne} is terminated when the one-step lookahead criterion in~\eqref{eqn:one:step:lookahead} is less than~$\num{5e-2}$. 

First, we compare Algorithm~\ref{algo:ne} with GPGame~\cite{PBH18}, which can used for `black box' utility functions of general games. GPGame requires an initial set of `space filling' measurements. As suggested in~\cite{PBH18}, we set the number of initial measurements to be~$11$. To enable comparison, we initialized Algorithm~\ref{algo:ne} with the same initial conditions. Table~\ref{tab:comparison:iterations} compares the number of iterations required by Algorithm~\ref{algo:ne} and GPGame. In Table~\ref{tab:comparison:iterations}, we used `Probability of Nash Equilibrium' defined in~\cite{PBH18} as the criterion for GPGame. Algorithm~\ref{algo:ne} performs better than GPGame in this setup. Algorithm~\ref{algo:ne} utilizes the structure of the potential game, i.e.\ the optimization of the potential function leads to Nash equilibrium. In addition, GPGame requires estimating $\nplayers$~negative quadrant probabilities of multi-variate Gaussians of size $|\actionset_\playeridx|$, which is computationally expensive. In comparison, estimating the one-step lookahead criterion in~\eqref{eqn:one:step:lookahead} is analytic (refer to~\eqref{eqn:Z:plus:dist}). 

Finally, we also compute the equilibrium through the exponential weight algorithm, under the \emph{no-regret learning} framework. The exponential weights algorithm is shown to work in potential games, even when only the measurements of the utility function are available, using the \emph{bandit estimator} defined in~\cite{HCM17}. 
However, due to the absence of a central agent (system operator), the exponential weights algorithm differs from the setting considered in this paper; see Figure~\ref{fig:blk}. 
Table~\ref{tab:comparison:iterations} shows the comparison of the number of iterations between Algorithm~\ref{algo:ne} and the exponential weight algorithm in~\cite{HCM17}. In comparison to the exponential weight algorithm, Algorithm~\ref{algo:ne} requires only a fraction of the number of iterations. However, the exponential weight algorithm has the advantage that it is fully distributed.  
\begin{table}[!h]
	\centering
	\begin{tabular}{|c|c|}
		\toprule
		Algorithm & \# Iterations\\
		\midrule
		Algorithm~\ref{algo:ne} &  2-3 \\
		GPGame~\cite{HCM17} &  4-5 \\
		Exponential weight~\cite{HCM17} &  200 \\
		\bottomrule
	\end{tabular}
	\caption{\footnotesize Comparison between the number of iterations between the various algorithms for computing Nash equilibria in finite potential games. In Algorithm~\ref{algo:ne} and GPGame, the number of initial `space filling' measurements was set to~$11$. Algorithm~\ref{algo:ne} performs better than GPGame, and uses only a fraction of iterations compared to a completely distributed algorithm like the exponential weight algorithm.  \label{tab:comparison:iterations} }
\end{table}

\textbf{Infinite Game Formulation: }The experiment was repeated with the formulation of continuous action sets in Sec.~\ref{sec:gp:continous:action:sets}. To run Algorithm~\ref{algo:ne:continous:adaptive}, the following parameters were chosen: 
\begin{equation}
  \begin{gathered}
    c_1 = \num{1e-4},\quad c_2 = 0.8,\quad c_w = 0.3, \quad \lambda_\playeridx^2 = \sqrt{30}, \\
\ell = 1, \quad \stepsize_{\text{max}} = 1, \quad \beta = 0.75
\end{gathered}
  \label{eqn:params:cournot:algo}
\end{equation}

\begin{figure}[!t]
  \centering
  \includegraphics[width=0.49\textwidth]{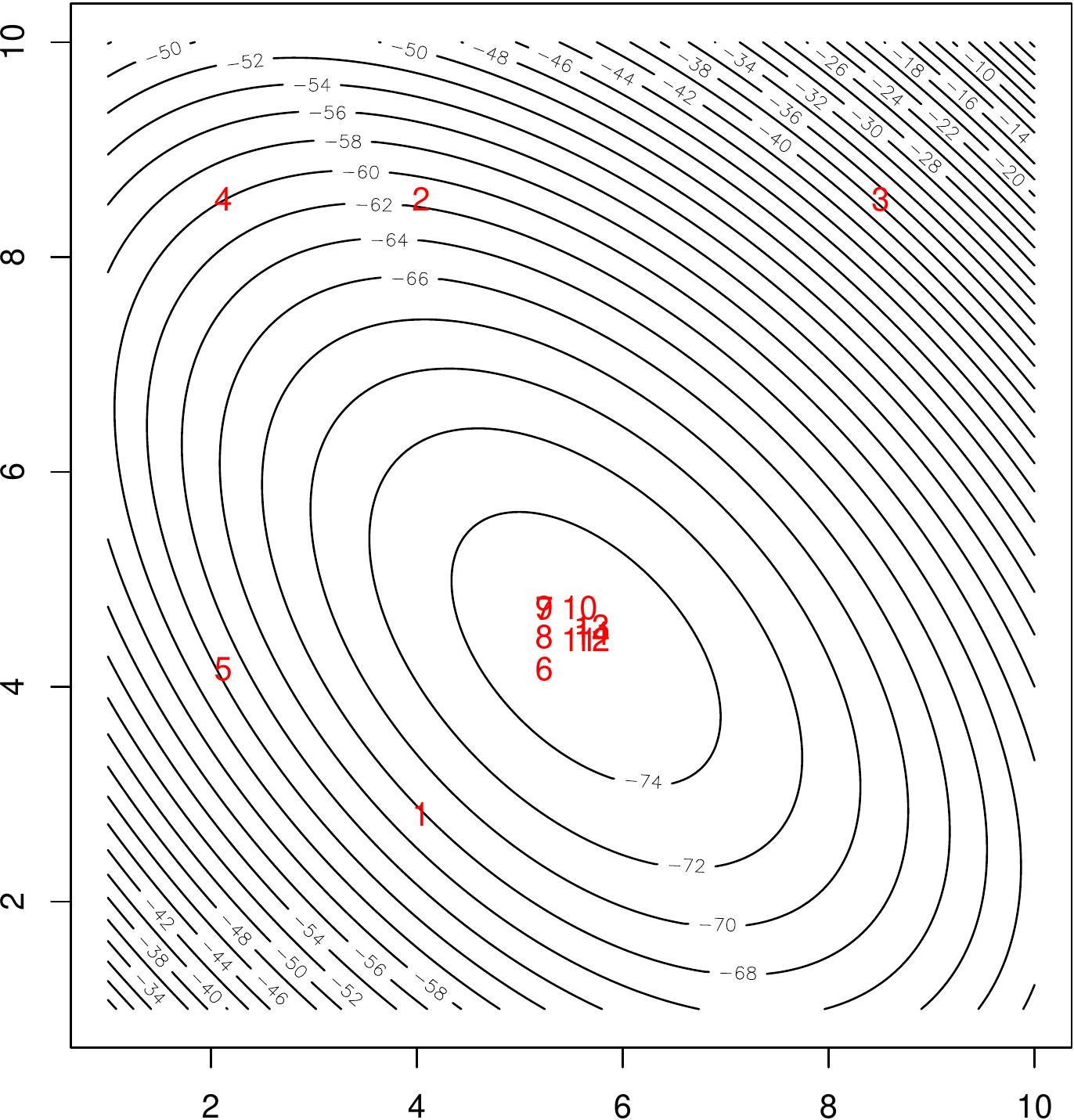}\llap{\raisebox{3.5ex}{
\begin{tikzpicture}
\begin{axis}[
width=1in,
height=1in,
at={(0,0)},
scale only axis,
xmin=1,
xmax=13,
ymin=-0.5,
ymax=6.5,
axis background/.style={fill=white},
xlabel={\footnotesize Iteration},
xlabel style = {yshift = 0.5em},
axis on top,
xmajorgrids,
xminorgrids,
tick align=inside,
xtick={1,3,...,13},
ytick={1,...,6.5},
ticklabel style = {font=\tiny},
ymajorgrids,
yminorgrids,
x tick label style={yshift = 1.75ex, xshift = 0.5ex},
y tick label style={xshift = 1.75ex, yshift = 0.5ex},
grid style={dotted}
]
\addplot[color=black, mark=*]
  table[row sep=crcr]{
  1 	5.97148891\\
  2 	4.46330517\\
  3 	6.38516367\\
  4 	4.56098205\\
  5 	3.11721225\\
  6 	0.58526927\\
  7 	0.27296727\\
  8 	0.28352693\\
  9 	0.25580781\\
  10 	0.31467793\\
  11 	0.11985216\\
  12 	0.13252376\\
  13 	0.07312566\\
};
\end{axis}
\end{tikzpicture}
  }}
  \caption{\footnotesize Convergence of Algorithm~\ref{algo:ne:continous:adaptive} for the Cournot oligopoly problem in~\eqref{eqn:params:cournot}. The path taken by the algorithm is shown in~{red}. The path is superimposed on the contour plot of the potential function (unknown to Algorithm~\ref{algo:ne:continous:adaptive}). It is easy to see that Algorithm~\ref{algo:ne:continous:adaptive} converges to the optimal solution. It can be noticed (from inset) that the algorithm initially selects larger step sizes as it explores the search space followed by smaller step sizes as it exploits the available information to reach the global optimum.\label{fig:fip:path:cont}
}
\end{figure}
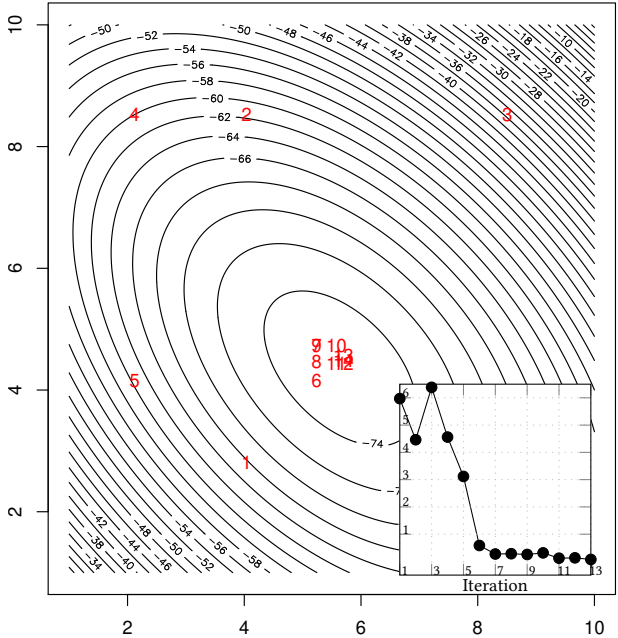

Figure~\ref{fig:fip:path:cont} shows the convergence of the Algorithm~\ref{algo:ne:continous:adaptive}. The algorithm is terminated when the one-step lookahead criterion less than $\num{1e-4}$. The potential function for the Cournot example is analytic and is given in~\cite{MonShap96}. In Figure~\ref{fig:fip:path:cont}, we also plot the contour plot of the potential function. From the inset in Fig.~\ref{fig:fip:path:cont} it is easy to see that the algorithm converges to the optimal solution. Also, notice that the algorithm initially selects larger step sizes as it explores the search space followed by smaller step sizes as it exploits the available information to reach the global optimum, a property of the one-step lookahead criterion in~\eqref{eqn:choose:player:infinite}. The GPGame algorithm in~\cite{PBH18} cannot be used when the action sets are continuous\footnote{We expect to extend some of the ideas in this paper to handle continuous action sets in GPGame.}. 

To further illustrate the efficiency of Algorithm~\ref{algo:ne:continous:adaptive}, we compared Algorithm~\ref{algo:ne:continous:adaptive} with a non-linear optimizer. For the non-linear optimizer, we used BOBYQA\footnote{BOBYQA performed the best among all the non-linear optimizers we tried.}~\cite{BOBYQA09}, an algorithm that uses a quadratic approximation of the objective function. The potential function for the Cournot example is given in~\cite{MonShap96}. Table~\ref{tab:comparison:iterations:BOBYQA} shows the comparison of the number of iterations averaged over $10$ independent runs with varying start points. Both the non-linear optimizer and Algorithm~\ref{algo:ne:continous:adaptive} were provided with the same start points. Algorithm~\ref{algo:ne:continous:adaptive} is able to achieve similar performance to a non-linear optimizer with full knowledge of the objective (the potential function). 
\begin{table}[!h]
	\centering
	\caption{\footnotesize Comparison between the number of iterations of Algorithm~\ref{algo:ne:continous:adaptive} with a non-linear optimizer. Algorithm~\ref{algo:ne:continous:adaptive} is able to achieve similar performance to a non-linear optimizer with full knowledge of the potential function.  \label{tab:comparison:iterations:BOBYQA} }
	\begin{tabular}{|c|c|}
		\toprule
		Algorithm & \# Iterations\\
		\midrule
		Non-linear optimizer &  9 \\
		(BOBYQA~\cite{BOBYQA09}) &  \\
		Algorithm~\ref{algo:ne:continous:adaptive} &  12 \\
		\bottomrule
	\end{tabular}
\end{table}

\subsection{Potential Differential Games: Computing Linear Nash equilibrium}
In Section~\ref{sec:numerical:results:static:games}, we illustrated the advantage of Bayesian optimization approach to static potential games. In this section, we consider dynamic potential games, in particular, dynamic potential games evolving in continuous times, referred to as differential games\footnote{Similar techniques also apply to discrete time dynamic potential games.}. Differential games model a wide variety of interactions in economics, finance, sociology and biology; see for example~\cite{Basar99}. In static potential games several techniques exist to find the potential function given the utility functions of the players. However, in differential potential games, and in general for dynamic games, computing the potential function is non-trivial, even with the knowledge of the utility function. 

To illustrate the main results, we consider the following classical differential game of a common pool of resource exploited by heterogeneous players~\cite{Long99}: 
\begin{equation}
  \begin{aligned}
    \dot{\statediffgame}(t) &= a \statediffgame(t) - \sum_i \action_i(t), \quad a \ge 0\\
	\statediffgame(0) &= \statediffgame_0.
  \end{aligned}
  \label{eqn:diff:game}
\end{equation}
The state~$\statediffgame$ in~\eqref{eqn:diff:game} represents the stock of the resource and~${a > 0}$ implies that the resource is renewable such as in fishing or logging in forestry. 
The utility of player~$\playeridx$ when the game is played for a horizon of time $T$ is given by:
\begin{equation}
  J_\playeridx(\action_\playeridx, \action_{-\playeridx}) = \int_0^T u_i(\statediffgame(t), \action(t)) \exp{(-\theta_\playeridx t)} \; dt, \quad \theta_\playeridx \ge a
  \label{eqn:diffgame:utility}
\end{equation}
where, $\action(t) = \left(\action_1(t), \action_2(t),\cdots,\action_\nplayers(t) \right)$ is the action of all the players at time~$t$ and~$u_i(\cdot)$ is the instantaneous utility of each player as a function of the state and the action of all the players. The parameter~${\theta_\playeridx}$ represents the discount rate of player~$\playeridx$.  
The Nash equilibrium for a differential game is defined similarly to~\eqref{eqn:Nash:equilibrium} as:  
\begin{equation}
  \action^*_{\playeridx} = \underset{\action_{\playeridx} }{\argmin\;} J_{\playeridx}(\action_\playeridx,\action^*_{-\playeridx}) \quad \forall \playeridx.
  \label{eqn:open:loop:nash}
\end{equation}
In the following, as a special case, we consider the following instantaneous utility function: 
\begin{equation}
  \utility_i(\statediffgame(t), \action(t)) =  \left( \action_\playeridx(t) \right)^{\alpha_\playeridx}, \quad \alpha_\playeridx > 0
  \label{eqn:diff:game:utility:potential}
\end{equation}
The parameter $\alpha_\playeridx$ in~\eqref{eqn:diff:game:utility:potential} captures the classical economic notion of `\emph{returns to scale}', i.e.\  the rate of increase in payoff (or utility) relative to investment (the action, in this case) . Industries such as mining have an increasing returns to scale (${\alpha_\playeridx > 1}$), while computer technology have a decreasing returns to scale (${\alpha_\playeridx \le 1}$).  
The structure of~\eqref{eqn:diff:game:utility:potential} ensures that~\eqref{eqn:diff:game} is a potential game. The above model also describe dynamics of single capital stock~\cite{dockner2014markov} and optimal dynamic scheduling for a common resource~\cite{ABEJW06}. 

\textbf{Linear Control Strategies: } In this paper, we focus on linear strategies, i.e.\ strategies of the form~$\action_\playeridx(t) = \gamma_\playeridx \statediffgame(t)$. The focus on linear strategies is primarily motivated by simplicity. In addition, linear strategies are known to be optimal when the state transition in~\eqref{eqn:diff:game} is linear in state and actions and the instantaneous objective function is homogeneous like in~\eqref{eqn:diff:game:utility:potential}~\cite{Long98}. However, the focus on linear strategies precludes some non-linear Nash control strategies. 

\begin{figure}[t]
  \centering
  \input{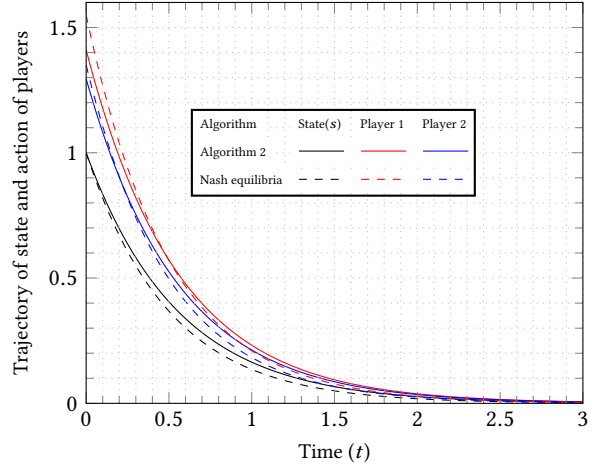}
  \caption{\footnotesize Comparison of the state and the linear control strategy obtained by Algorithm~\ref{algo:ne:continous:adaptive} against Nash equilibrium.}
  \label{fig:linear:strategy}
\end{figure}

\textbf{Simulation Setup and Results}:  
To illustrate the main results, we consider a version of the problem in~\eqref{eqn:diff:game} with $2$~players, with the following parameters:
\begin{equation}
  \begin{gathered}
  a = 0.9, \quad s_0 = 1, \quad \alpha_1 = 0.3, \quad \alpha_2 = 0.2, \\
  \theta_1 = \theta_2  = 0.95,\quad T = 4.
\end{gathered}
  \label{eqn:diff:game:parameters}
\end{equation}
Adopting the linear strategy, the potential function is parameterized by~$\gamma_1$ and~$\gamma_2$. 
For the linear strategy, the state evolution is exponential with parameter $a - \sum_\playeridx \gamma_\playeridx$. The utilities in~\eqref{eqn:diffgame:utility} were computed using numerical integration with $\num{1e4}$~integration points. Algorithm~\ref{algo:ne:continous:adaptive} was run with the parameters given in~\eqref{eqn:params:cournot:algo}, except for the hyperparameters $\lambda_j = \sqrt{7}$.  
Figure~\ref{fig:linear:strategy} shows the trajectory of the state and the linear strategy (actions of the players) after~$18$ iterations of Algorithm~\ref{algo:ne:continous:adaptive}. The `true' Nash equilibrium for the problem in~\eqref{eqn:diff:game} can be computed using the methodology outlined in~\cite{Long99}. From Fig.~\ref{fig:linear:strategy}, it can be observed that the strategy computed from Algorithm~\ref{algo:ne:continous:adaptive} is `close' to the Nash equilibrium.  
\label{sec:differential:games}
\subsection{Discussion}
In Sec.~\ref{sec:numerical:results:static:games} and Sec.~\ref{sec:differential:games}, we illustrated the advantage of Bayesian optimization approach to computing Nash equilibria of `black box' potential games. Even in the absence of knowledge of the utility function, the Nash equilibrium can be computed quite efficiently. However, this efficiency comes at a price. Algorithm~\ref{algo:ne} and Algorithm~\ref{algo:ne:continous:adaptive} require inverting a $K \times K$~matrix, which requires~$\mathcal{O}(K^3)$ computations. Sparse Gaussian process techniques~\cite{QR05} reduce the number of computations to~$\mathcal{O}(m^2K)$, where $m \ll K$. In addition, Algorithm~\ref{algo:ne:continous:adaptive}, requires computing the integrals in~\eqref{eqn:potential:covariance} and~\eqref{eqn:eta:covariance}. For the squared exponential kernel in~\eqref{eqn:mu:kernel:choice}, the $\eta$ and $\gamma$ integrands are analytic and expressions can be found in~\cite{HK13}. However, computing $\pi$ in~\eqref{eqn:potential:covariance} requires numerical integration. This could be computationally expensive as the number of players increase.  
\section{Conclusion \& Future Work}
\label{sec:conclusion}
In this paper, we considered potential games with `black box utility' functions. 
Using the Gaussian process framework and structure of potential games, we derived two novel algorithms; one for discrete action sets, and one for action sets with real intervals. We illustrated the efficiency of the algorithms, in terms of black box evaluations, for computing the Nash equilibrium of static games and the linear Nash equilibria of differential games. In addition, the algorithms provide a general non-parametric technique to compute Nash equilibria of potential games, without explicitly computing the potential function. 

Extensions of the current work could involve developing algorithms for computing the Nash equilibrium of population games, investigating sparse Gaussian process for reducing the computational complexity and computing closed loop Nash equilibrium of dynamic potential games. 
These issues promise to offer interesting avenues for future work.

\end{document}